\documentclass{ceurart}

\usepackage{float}
\restylefloat{table}

\graphicspath{{fig/}}

\begin{document}

\copyrightyear{2021}
\copyrightclause{Copyright for this paper by its authors.\\
  Use permitted under Creative Commons License Attribution 4.0
  International (CC BY 4.0).}

\conference{}
\title{When no news is bad news - Detection of negative events from news media content}

\author[1,2]{Kristoffer L. Nielbo}[%
orcid=0000-0002-5116-5070,
]
\ead{kln@cas.au.dk}
\ead[url]{https://knielbo.github.io/}

\author[1]{Frida Haestrup}[
orcid=0000-0001-7561-6297,
]
\ead{frihae@cas.au.dk}

\author[1]{Kenneth C. Enevoldsen}[
orcid=0000-0001-8733-0966,
]
\ead{kenneth.enevoldsen@cas.au.dk}

\author[1]{Peter B. Vahlstrup}[
orcid=0000-0003-4126-7740,
]
\ead{imvpbv@cc.au.dk}

\author[2]{Rebekah B. Baglini}[
orcid=0000-0002-2836-5867,
]
\ead{rbkh@cc.au.dk}

\author[2]{Andreas Roepstorff}[
orcid=0000-0002-3665-1708,
]

\address[1]{Center for Humanities Computing Aarhus, Jens Chr. Skous Vej 4, Building 1483, 3rd floor, DK-8000 Aarhus C, Denmark}

\address[2]{Interacting Minds Centre, Jens Chr. Skous Vej 4, Building 1483, 3rd floor, DK-8000 Aarhus C, Denmark}

\begin{abstract}
During the first wave of Covid-19 information decoupling could be observed in the flow of news media content. The corollary of the content alignment within and between news sources experienced by readers (i.e., all news transformed into Corona-news), was that the novelty of news content went down as media focused monotonically on the pandemic event. This all-important Covid-19 news theme turned out to be quite persistent as the pandemic continued, resulting in the, from a news media's perspective, paradoxical situation where the same news was repeated over and over. This information phenomenon, where novelty decreases and persistence increases, has previously been used to track change in news media, but in this study we specifically test the claim that new information decoupling behavior of media can be used to reliably detect change in news media content originating in a \emph{negative event}, using a Bayesian approach to change point detection.
\end{abstract}


\maketitle

\section*{Introduction}
A peculiar behavior could be observed in news media when the first wave of Covid-19 virus spread across the world. In response to this pandemic event, the ordinary rate of change in news content was disrupted because every story became associated with Covid-19. On the one hand, content novelty went down, because every story became more similar to previous stories, but on the other hand, the Covid-19 association became more prevalent, resulting in, at least initially, an increase in content persistence. A recent study \cite{nielbo_news_2021} argues that this behavior is an example of the \emph{news information decoupling} (NID) principle, according to which information dynamics of news media are (initially) decoupled by temporally extended catastrophes such that the content novelty decreases as media focus monotonically on the catastrophic event, but the resonant property of said content increases as its continued relevance propagate throughout the news information system. The authors further argued NID can be used to detect significant change in news media that originate in catastrophic events.

Previous studies have shown that variation in newspapers' word usage is sensitive to the dynamics of socio-cultural events \cite{guldi_measures_2019, van_eijnatten_eurocentric_2019, daems_workers_2019}, can detect event-driven shifts \cite{kestemont_mining_2014}, and accurately can model effects of change in comprehensive collections of newspapers \cite{bos_quantifying_2016}. Furthermore, the associative structure of newspapers has been shown to accurately capture thematic development \cite{newman_probabilistic_2006}, and, when modelled dynamically, is indicative of the evolution of cultural values and biases \cite{van_eijnatten_eurocentric_2019, wevers_using_2019}. Adaptive fractal analysis of word frequencies over time has been used to discriminate between different classes of catastrophic events that display class-specific fractal signatures in, among other things, word usage in newspapers \cite{gao_culturomics_2012}. Several studies have shown that information theoretical construct can be used to detect fundamental conceptual differences between distinct periods \cite{guldi_measures_2019},  concurrent normative and ideological movements \cite{barron_individuals_2018}, and even, development of ideational factors (e.g., creative expression) in temporally dependent writings \cite{murdock_exploration_2015, nielbo_automated_2019, nielbo_curious_2019}. More specifically, a set of methodologically related studies studies have applied windowed relative entropy to thematic text representations to generate signals that capture information \emph{novelty} as a reliable content difference from the past and \emph{resonance} as the degree to which future information conforms to said novelty \cite{barron_individuals_2018, murdock_exploration_2015}. Three recent studies have found that successful social media content show a strong association between novelty and resonance \cite{nielbo_trend_2021}, that legacy news media under normal conditions display a remarkably similar medium to strong association between novelty and resonance across the political spectrum \cite{nielbo_news_2021}, and, finally, that variation in the novelty-resonance association can predict significant change points in historical data \cite{vrangbaek_composition_2021}.

This study specifically tests the claim of \cite{nielbo_news_2021} that NID-like behavior can provide input for change point detection algorithms. Specifically, we propose to test the claim that two change points are observable in news media during the first phase of Covid-19, $Lockdown$ and $Opening$ respectively, using a Bayesian approach to change point detection.

\begin{figure}
	\centering
	    \includegraphics[width=1\textwidth]{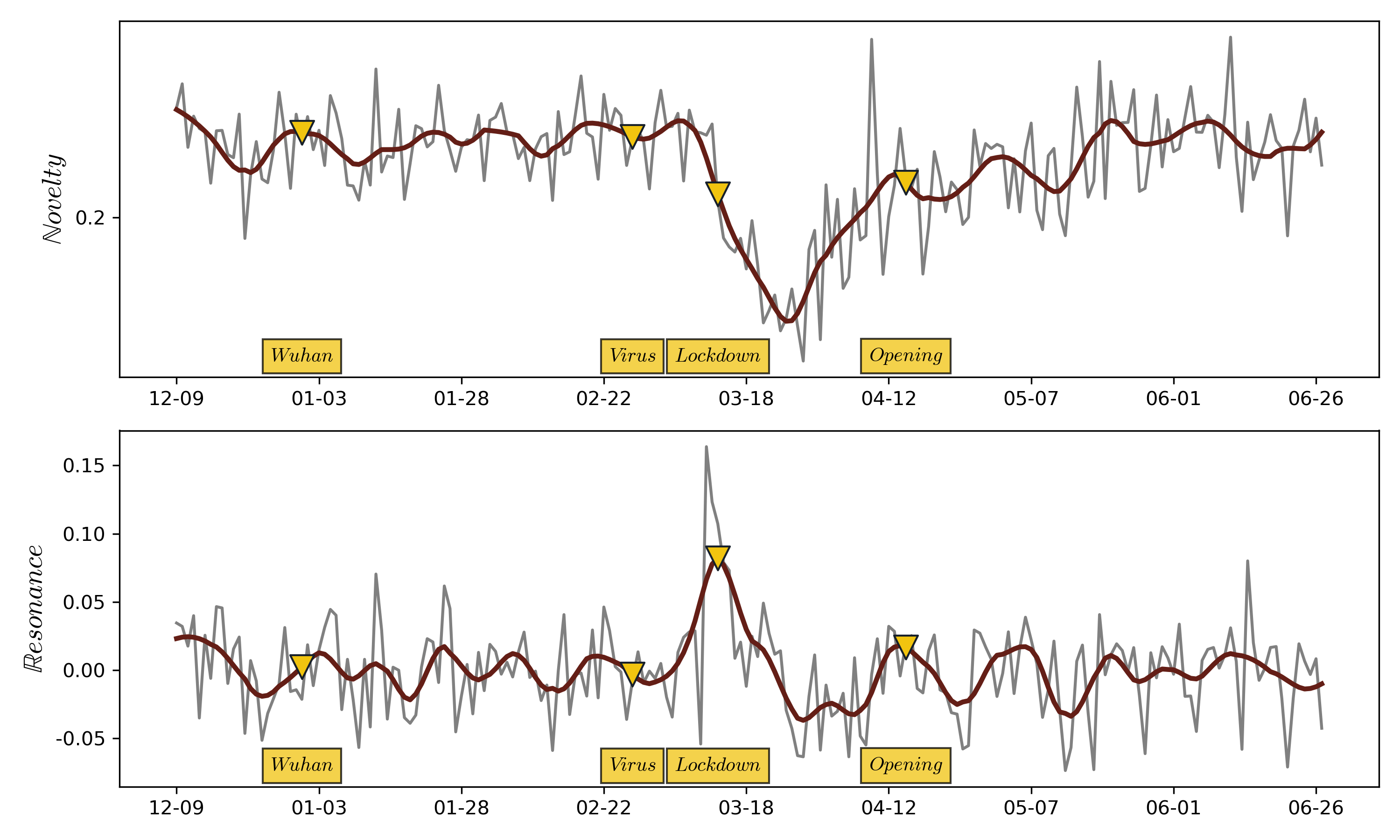}\\
\caption{Novelty (upper panel) and resonance (lowerpanel) for the center-left newspaper \textit{Politiken} before and during Covid-19 phase 1. Trend lines in the upper and middle panel are estimated using a nonlinear adaptive filter suggested in \cite{nielbo_news_2021}.}
    \label{fig:poladapt}
\end{figure}

\section*{Results}

Figure \ref{fig:poladapt} displays a prototypical example of NID during the first phase of Covid-19 \cite{nielbo_news_2021}. Although Covid-19 news items date back to December 2019, `$Wuhan$', newspaper content is not impacted until the period after the first national outbreak (in this case in Denmark). `$Virus$`. From the phase 1 lockdown `$Lockdown$` to the opening, `$Opening$`, the newspaper shows a valley in novelty and, initially, a peak in resonance until both processes approximately return to normal after the opening.

\begin{table}[H]
    \centering
    \begin{tabular}{lcccc}
    \hline
    Source & Class & \textit{NID} Start & \textit{NID} End & \textit{NID}\\
    \hline
        Berlingske & $B$ & $03.07~[03.03, 03.09]$ & $04.28~[04.09, 05.08]$ & $True$ \\
        BT & $T$ & $04.10~[12.30, 09.01]$ & $07.25~[04.22, 09.03]$ & $False$ \\
        Ekstrabladet & $T$ & $01.28~[01.02, 03.17]$ & $05.08~[01.16, 07.22]$ & $False$ \\
        Jyllands-Posten & $B$ & $03.10~[03.08, 03.14]$ & $05.25~[05.21, 06.06]$ & $True$ \\
        Kristligt Dagblad & $B$ & $03.07~[03.05, 03.12]$ & $04.15~[04.11, 04.17]$ & $True$\\
        Politiken & $B$ & $03.13~[03.12, 03.13]$ & $04.08~[04.05, 04.08]$ & $True$\\
        \hline
    \end{tabular}
    \caption{Estimated temporal change points at $94\%$ high density intervals for novelty. Column one contains the name of the newspaper, columns two its class ($B$roadsheet or $T$abloid), NID Start and End is the beginning and end of the lockdown as represented in the newspaper, and the final column indicated if the specific source supported the NID principle.}
    \label{tab:table1}
\end{table}

\begin{figure}
    \centering
    \includegraphics[width=\textwidth]{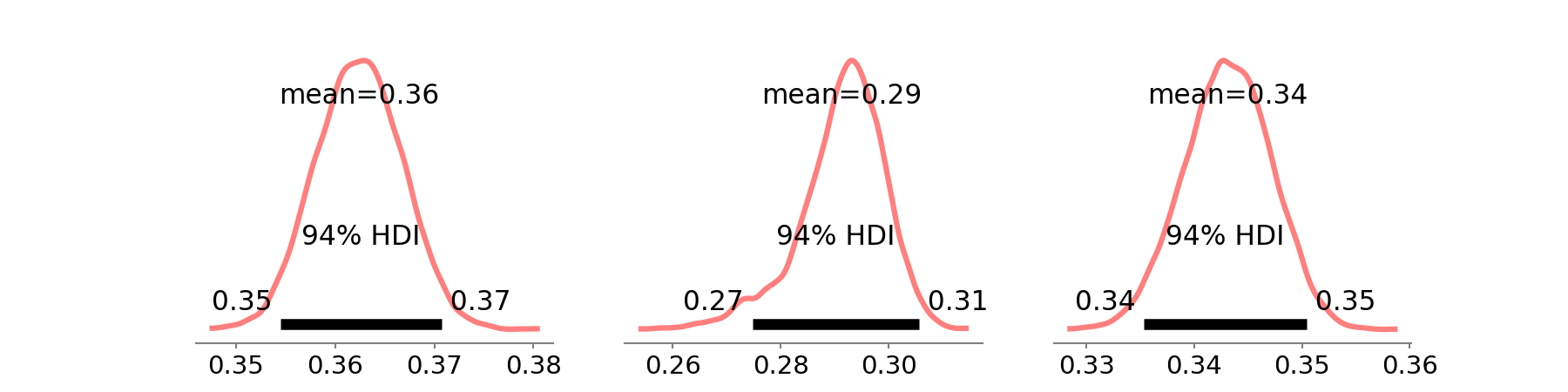}\\
    \includegraphics[width=\textwidth]{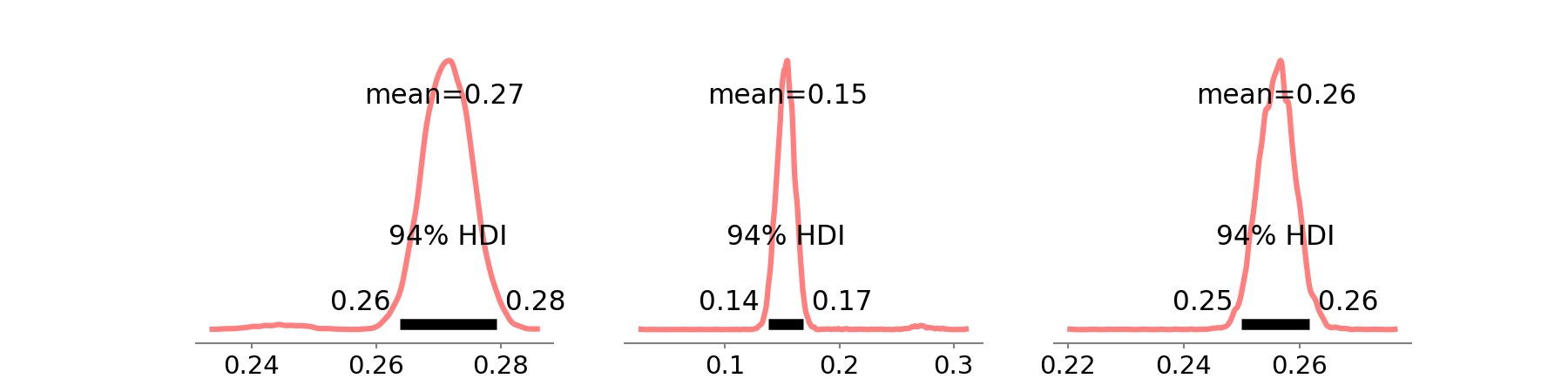}
    \caption{Posterior distributions of novelty at $94\%$ high density intervals for newspapers Berlingske, and Politiken, see table \ref{tab:table2}.}
    \label{fig:posterior}
\end{figure}

To validate the observed behavior, we tested for two change points in novelty using a Bayesian model, see Appendix \ref{appA} for methods. The first change point, `$NID$ Start` should separate pre-lockdown from lockdown centered on week 11, and the second lockdown, `$NID$ End` from post opening (centered on week 16). Table \ref{tab:table1} shows the estimated change points for six national newspapers, two of which are $T$abloid newspapers (Class) and the remainder $B$roadsheet. From the model, it can be observed that all broadsheet newspapers seem to support the NID principle in novelty. The first change point is placed in weeks 10-11, the second, however, is more a matter of contention. The opening change point lies within April and displays a month's delayed response. Finally, it can be observed that tabloid press shows no indication of NID behavior. Table \ref{tab:table2} and figure \ref{fig:posterior} show the posterior distributions, their means and highest density intervals, for four broadsheet and one tabloid newspaper, clearly indicating that broadsheet newspapers do conform to NID, while tabloids do not. 

\begin{table}[H]
    \centering
    \begin{tabular}{lccc}
    \hline
    Source & $\mathbb{N}_{pre}$ & $\mathbb{N}_{NID}$ & $\mathbb{N}_{post}$\\
    \hline
        Berlingske &  $0.36~[0.35, 0.37]$ &  $0.29~[0.27, 0.31]$ &  $0.34~[0.34, 0.35]$ \\
        Jyllands-Posten & $0.29~[0.28, 0.30]$ &  $0.23~[0.22, 0.24]$ & $0.27~[0.26, 0.28]$ \\
        Kristligt Dagblad & $0.27~[0.26, 0.28]$ & $0.19~[0.18, 0.21]$ & $0.26~[0.25, 0.27]$ \\
        Politiken & $0.27~[0.26, 0.28]$ & $0.15~[0.14, 0.17]$ & $0.26~[0.25, 0.26]$ \\
        \hline
    \end{tabular}
    \caption{Novelty values at $94\%$ high density intervals before during and after the lockdown for the four broadsheet newspapers that supported the NID principle, see table \ref{tab:table1}.}
    \label{tab:table2}
\end{table}

\begin{figure}
    \centering
    \includegraphics[width=.32\textwidth]{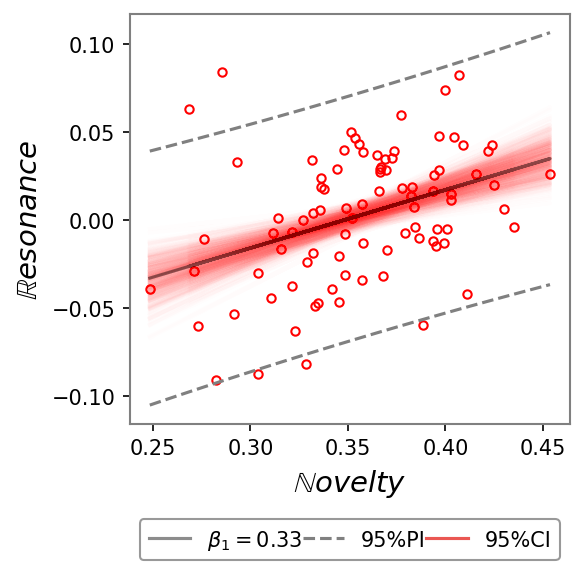}
    \includegraphics[width=.32\textwidth]{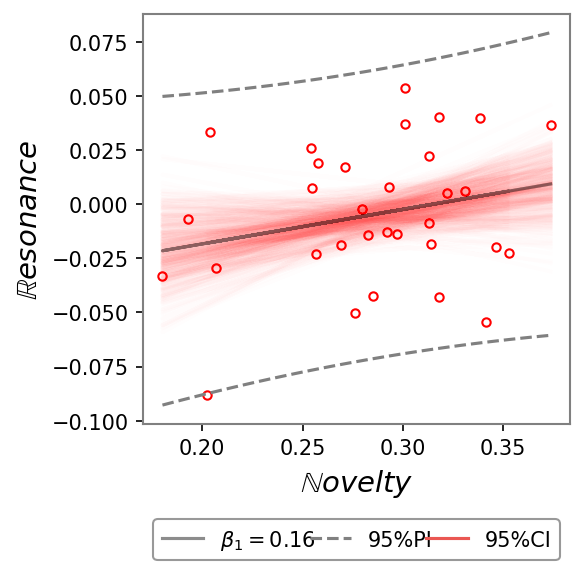}
    \includegraphics[width=.32\textwidth]{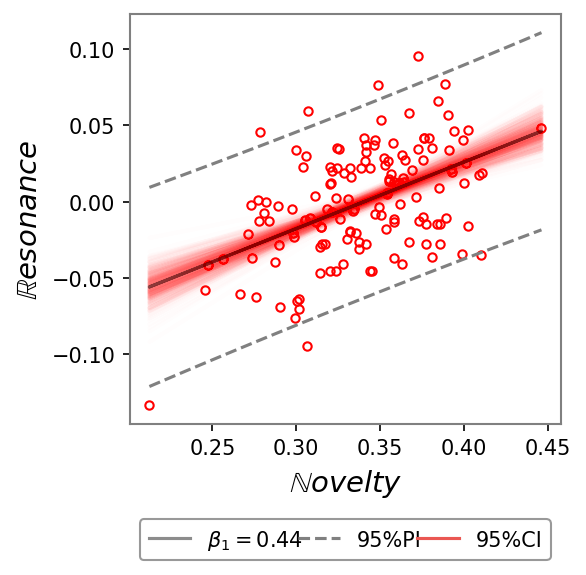}\\
    
    \includegraphics[width=.32\textwidth]{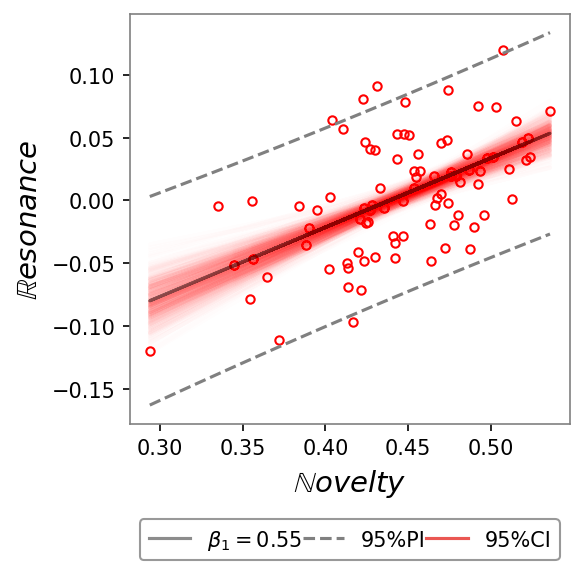}
    \includegraphics[width=.32\textwidth]{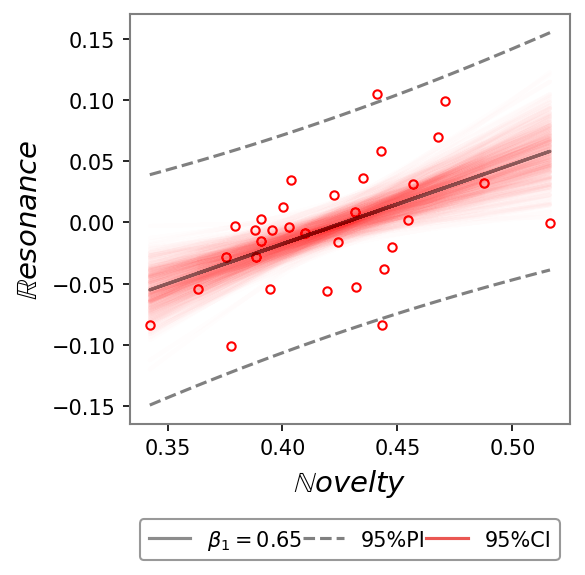}
    \includegraphics[width=.32\textwidth]{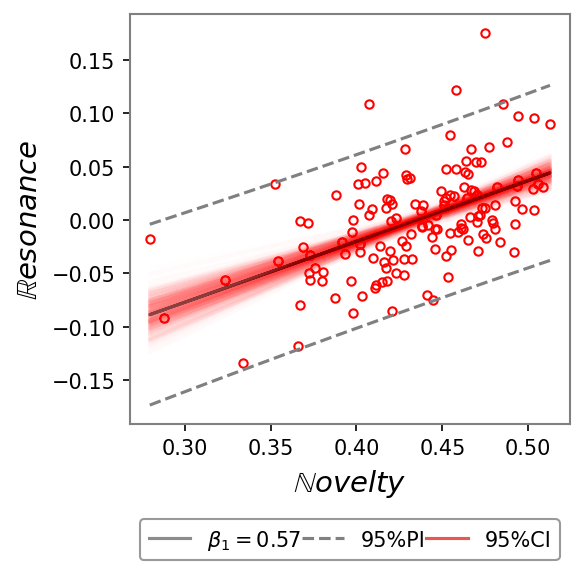}\\
    
    \includegraphics[width=.32\textwidth]{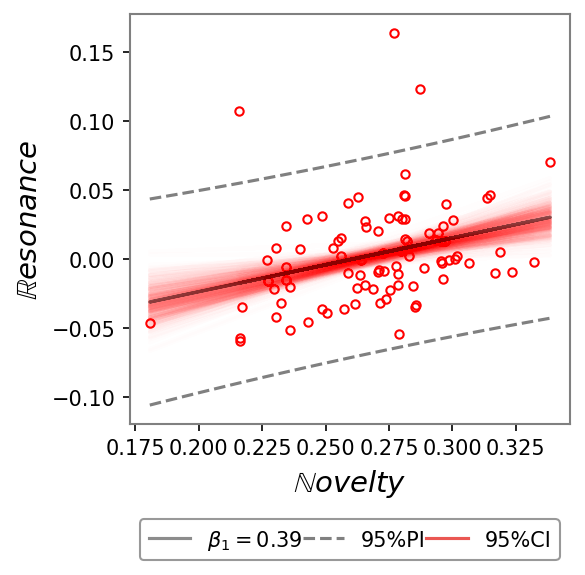}
    \includegraphics[width=.32\textwidth]{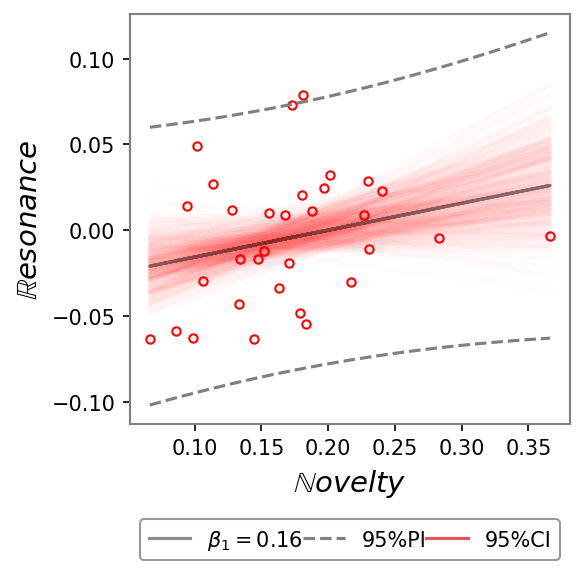}
    \includegraphics[width=.32\textwidth]{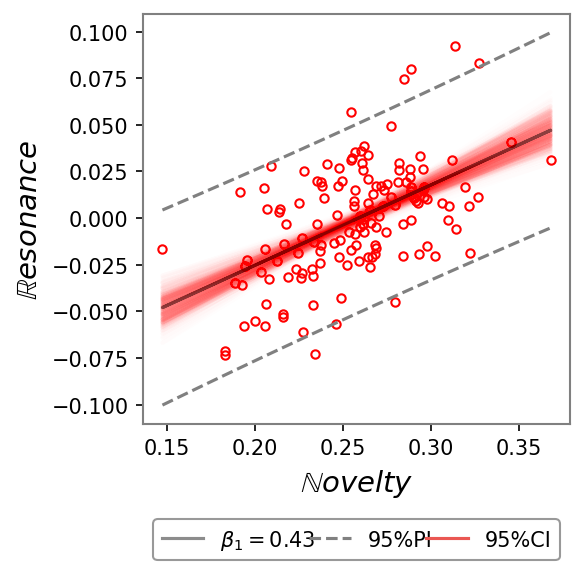}
    
    \caption{$\mathbb{N} \times \mathbb{R}$ slopes before during and after the lockdown for Berlingske (upper row), Ekstrabladet (middle row), and Politiken (lower row) during Covid-19 phase 1.}
    \label{fig:resonance}
\end{figure}

That novelty decreases during a catastrophic event is nevertheless only half the story. For NID to be supported by the data, resonance should increase during the lockdown such that the medium to strong association between novelty and resonance is momentarily weakened. Following \cite{nielbo_news_2021}, we inspected the time-windowed linear fits of resonance on novelty, $\mathbb{N} \times \mathbb{R}$, in order to confirm this, see figure \ref{fig:resonance}.  All broadsheet newspapers display a slope decrease during the lockdown, thereby conforming to the NID principle \ref{tab:table3}. Tabloids on the other hand, follow an inverse pattern, such that the $\mathbb{N} \times \mathbb{R}$ slope increases during the lockdown period.

\begin{table}[H]
    \centering
    \begin{tabular}{lccc}
    \hline
    Source & $\mathbb{N} \times \mathbb{R}_{pre}$ & $\mathbb{N} \times \mathbb{R}_{NID}$ & $\mathbb{N} \times \mathbb{R}_{post}$\\
    \hline
        Berlingske &  $0.33~[0.17, 0.51]$ &  $0.16~[-0.07, 0.38]$ &  $0.44~[0.32, 0.58]$\\
        BT &  $0.49~[0.29, 0.66]$ &  $0.55~[0.28, 0.83]$ &  $0.26~[0.08, 0.43]$\\
        Ekstrabladet &  $0.55~[0.38, 0.72]$ &  $0.65~[0.26, 1]$ &  $0.57~[42, 0.71]$\\
        Jyllands-Posten &  $0.42~[0.24, 0.63]$ &  $0.31~[0.04, 0.56]$ &  $0.39~[0.25, 0.51]$\\
        Kristligt Dagblad &  $0.57~[0.34, 0.78]$ &  $0.43~[0.06, 0.78]$ &  $0.76~[0.55, 0.95]$\\
        Politiken &  $0.39~[0.14, 0.61]$ &  $0.16~[-0.05, 0.37]$ &  $0.43~[0.32, 0.54]$\\
        \hline
    \end{tabular}
    \caption{NxR coefficients at 95\% confidence intervals before during and after the lockdown for all newspapers in the sample.}
    \label{tab:table3}
\end{table}

\section*{Concluding Remarks}
In conclusion, this study sought to validate the news information decoupling (NID) principle on a sample of six national Danish newspapers during the first phase of Covid-19. Using a Bayesian approach to change point detection, we found that content novelty in broadsheet newspapers does indeed display statistically reliable points of change during the Covid-19 lockdown. NID was further corroborated by the $\mathbb{N} \times \mathbb{R}_{pre}$ slopes that indicated a decoupling of resonance from novelty during the lockdown. Several observations can be made from the findings. First, the estimated change points for the `Pre-lockdown $\rightarrow$ Lockdown' are spread over a two week interval, which reflects that a lockdown could be reasonably predicted already from the first Covid-19 incident in Denmark. Second, in a similar vein the `Lockdown $\rightarrow$ Opening` change points are spread over an entire month from April 8 to May 8. The Danish government during the period was center-left and model's uncertainty in determining the opening may reflect political observations \cite{nielbo_news_2021}, where center-right newspapers (e.g., Berlingske and Jyllands-Postern) were more sceptical towards the government's implementation of an opening than the center-left (e.g., Politiken). In other words, the center-right might have been more reluctant to acknowledge the opening as a return to normal. Third, tabloid newspapers do not show any indication of a news decoupling, on the contrary, their $\mathbb{N} \times \mathbb{R}_{pre}$ slopes momentarily increases during the lockdown . This increase in slopes does, however, not provide much useful information, because, as shown by the change point detection model, the periodization is not meaningful to the two tabloid newspapers. 

Validation of the NID principle is still needed for multilingual data and its value for crisis management should be further tested. For change detection, the scope of the principle needs additional testing; does NID generalize beyond a small set of negative events to, for instance, temporally extended significant events (e.g., moon landing, fall of the Berlin Wall). Finally, several comparisons already hinted that left vs. right-wing newspapers, tabloid vs. broadsheet newspapers, silly season and other seasonal effects, are interesting venues for media and journalism researchers.    

\section*{Appendix}
    \appendix

\section{Methods}\label{appA}

\subsection*{Data and Normalization}
The data set consists of all linguistic content (title and body text) from front pages of six Danish national newspapers Berlingske, BT, Ekstrabladet, Jyllands-Posten, Kristligt Dagblad, and Politiken. The newspapers were sampled during December 1, 2019 to July 1 2020. Content not produced by the newspaper, e.g., advertisements, was excluded from the sample. In order to normalize linguistic content, numerals and highly frequent function words were removed, and the remaining data were lemmatized and casefolded. Subsequently, the data were represented as a bag-of-words (BoW) model using latent Dirichlet allocation in order to generate a dense low-rank representation of each article. Note that with a few modifications to equations \eqref{eq:4} and \eqref{eq:5}, the approach works for any probabilistic or geometric vector-representation of documents. Novelty and resonance were estimated for in windows of one week ($w = 7)$.

\subsection*{Novelty and Resonance}
Two related information signals were extracted from the temporally sorted BoW model: \textit{Novelty} as an article $s^{(j)}$'s reliable difference from past articles $s^{(j-1)}, s^{(j-2)} , \dots ,s^{(j-w)}$ in window $w$:

\begin{equation}
\mathbb{N}_w (j) = \frac{1}{w} \sum_{d=1}^{w}  JSD (s^{(j)} \mid s^{(j - d)})\label{eq:1}
\end{equation}

and \textit{resonance} as the degree to which future articles $s^{(j+1)}, s^{(j+2)}, \dots , s^{(j+w)}$ conforms to article $s^{(j)}$'s novelty:

\begin{equation}
\mathbb{R}_w (j) = \mathbb{N}_w (j) - \mathbb{T}_w (j)\label{eq:2}
\end{equation}

where $\mathbb{T}$ is the \textit{transience} of $s^{(j)}$:

\begin{equation}
\mathbb{T}_w (j) = \frac{1}{w} \sum_{d=1}^{w}  JSD (s^{(j)} \mid s^{(j + d)})\label{eq:3}
\end{equation}

The novelty-resonance model was originally proposed in \cite{barron_individuals_2018}, but here we propose a symmetrized and smooth version by using the Jensen–Shannon divergence ($JSD$):

\begin{equation}
JSD (s^{(j)} \mid s^{(k)}) =  \frac{1}{2} D (s^{(j)} \mid M) + \frac{1}{2} D (s^{(k)} \mid M)\label{eq:4}
\end{equation}

with $M = \frac{1}{2} (s^{(j)} + s^{(k)})$ and $D$ is the Kullback-Leibler divergence:

\begin{equation}
D (s^{(j)} \mid s^{(k)}) = \sum_{i = 1}^{K} s_i^{(j)} \times \log_2 \frac{s_i^{(j)}}{s_i^{(k)}}\label{eq:5}
\end{equation}

Finally, in order to describe the information states before and after an events (e.g., Lockdown, Opening), we fit resonance on novelty to estimate the $\mathbb{N}\times\mathbb{R}$ slope $\beta_1$ in the specific time windows:

\begin{equation}
\mathbb{R}_i = \beta_0 + \beta_1 \mathbb{N}_i + \epsilon_i, ~~ i = 1, \dots, n.\label{eq:7}
\end{equation}

\subsection*{Bayesian Change Point Detection}

For the estimation of change points, a Bayesian approach was used. Following previous considerations, we assume that the time series contains two change points, $\tau_1$ and $\tau_2$. Aside from change points, the series is assumed to be stable and follow a normal distribution with varied mean, $\mu_i$, and singular variance, $\sigma$. This gives us the following model given the observed Novelty, $\mathbb{N}_i$:


\begin{equation}
  \mathbb{N}_{t} =
    \begin{cases}
      \text{$\mathcal{N}(\mu_1, \sigma)  \text { for }  t<\tau_{1}  $ } \\
      \text{$\mathcal{N}(\mu_2, \sigma)  \text { for }  \tau_{1}\leq t <\tau_{2} $ } \\
      \text{$\mathcal{N}(\mu_3, \sigma)  \text { for }   t\geq\tau_{2}$ } \\
    \end{cases}       
\end{equation}

for which we wish to estimate the location of the change points $\tau_i$, means $\mu_i$ and variance $\sigma$, i.e. the following posterior:

$$P(\mu_i,  \sigma , \tau_i | \mathbb{N}_t) = P(\mu_1 , \mu_2, \mu_3, \sigma , \tau_1, \tau_2 | \mathbb{N}_t)$$

For estimation of the posterior, we have used NUTS sampling as implemented in pyMC3 \citep{pymc3} using 4000 samples. The estimation was done using using naive to slightly conservative priors assuming that the change points, $\tau_i$, can be anywhere in the sequence (with $\tau_2 > \tau_1$) and that the variance, $\sigma$, is stable across change points. Note that the half Cauchy prior distribution has series of beneficial properties \citep{gelman_prior_nodate, polson_half-cauchy_2012} including its fat tail which allows for extreme values. These assumptions were modelled using the following priors:

\begin{equation*}
    \begin{split}
        \mu_i & \sim \mathcal{N}(0, 0.5)                   \\
        \sigma & \sim \text{Half Cauchy}(0.5)               \\
        \tau_1 & \sim \text{Uniform}(0, \text{max}(\mathbb{N}_{t}))        \\
        \tau_2 & \sim \text{Uniform}(\tau_1, \text{max}(\mathbb{N}_{t}))   \\
    \end{split}
\end{equation*}

\section{Online Resources}

All data are proprietary and have been collected through Infomedia's API: \url{https://infomedia.dk/}. For inquiries regarding models and derived data, please contact \url{kln@cas.au.dk}. The source code for methods is available on Github: \url{https://bit.ly/3beahFd}. More details on NID detection can be found at NeiC's NDHL website: \url{https://bit.ly/3bfeW9C}.

\begin{acknowledgments}
This research was supported the "HOPE - How Democracies Cope with COVID-19"-project funded by The Carlsberg Foundation with grant CF20-0044, NeiC's Nordic Digital Humanities Laboratory project, and DeiC Type-1 HPC with project DeiC-AU1-L-000001. The authors would like to thank Berlingske Media, JP/Politkens Hus, and Kristeligt Dagblad for providing access to proprietary data.
\end{acknowledgments}

\bibliography{main}
\end{document}